\documentclass[12pt,a4paper,final]{iopart}

\usepackage{bm}
\usepackage{iopams}  
\usepackage{graphicx}
\usepackage[breaklinks=true,colorlinks=true,linkcolor=blue,urlcolor=blue,citecolor=blue]{hyperref}

\begin{document}

\title{Graphene Cantilever Under Casimir Force}

\author{Amel Derras-Chouk$^1$, Eugene M. Chudnovsky$^{1}$, Dmitry A. Garanin$^1$ and Reem Jaafar$^{2}$}
\address{$^1$Department of Physics and Astronomy, Lehman College and Graduate School, The City University of New York, 250 Bedford Park Boulevard West, Bronx, NY 10468-1589}
\ead{aderraschouk@gradcenter.cuny.edu  \vspace{1mm}}

\address{$^{2}$Department of Mathematics, Engineering and Computer Science, LaGuardia Community College, The City University of New York, 31-10 Thomson Avenue, Long Island City, NY 11101}

\begin{abstract}
Stability of graphene cantilever under Casimir attraction to an underlying conductor is investigated. The dependence of the instability threshold on temperature and flexural rigidity is obtained. Analytical work is supplemented by numerical computation of the critical temperature above which the graphene cantilever irreversibly bends down and attaches to the conductor. The geometry of the attachment and exfoliation of the graphene sheet is discussed. It is argued that graphene cantilever can be an excellent tool for precision measurements of the Casimir force. 
\end{abstract}

\section{Introduction}

When two electrically neutral bodies are separated by distances comparable to molecular sizes the interaction between them is usually referred to as a van der Waals force. It can be viewed within quantum theory (F. London \cite{London}) as a sum of electric dipolar forces between neutral atoms of the two bodies in close proximity to each other. Casimir \cite{Casimir} studied the force between two parallel conducting plates at separations greatly exceeding atomic distances by computing the energy of quantum fluctuations of the electromagnetic field in the free space between the plates. This force originates from the dependence of the zero-point energy of electromagnetic radiation on the boundary conditions provided by the surfaces of the interacting bodies. For parallel plates London's and Casimir's approaches have been reconciled by Lifshitz \cite{Lifshitz} within electrodynamics of continuous media that considers frequency and wave-vector dependent material susceptibilities. The problem has been intensively studied theoretically and experimentally in modern times, with a number of review articles written on the subject \cite{Kardar,Lamoreaux,Moststepanenko,Lambrecht}. A large body of work on Casimir and van der Waals forces has been done in recent years in application to graphene, see reviews, Refs.\  \cite{Klim,Woods}.

\begin{figure}[ht]
\vspace{1.2cm}
\begin{center}
\includegraphics[width=8.7cm,angle=0]{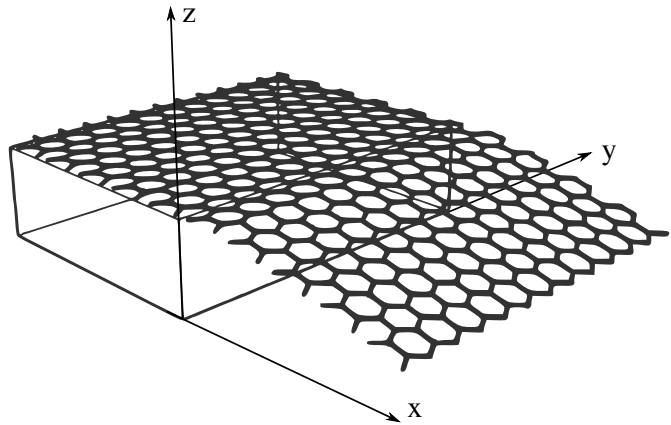}
\caption{Graphene cantilever above the surface of a conductor.}
\label{GC}
\end{center}
\end{figure}

Electromagnetic forces between electrically neutral bodies rapidly decrease on increasing separation. This explains why measurements of the Casimir force had to be performed at a micro- or nanoscopic scale \cite{Lamoreaux,Klim,Tang}. It also points towards the possibility that quantum and thermal fluctuations of the electromagnetic field can affect functionality of micro- and nano-electromechanical systems. Graphene stands out for future applications in such systems due to its unique mechanical, thermal, and electronic properties \cite{Neto,Novoselov}. Any design that involves suspended graphene would have to take into account Casimir interaction with surrounding elements. As a generic example we study in this paper the stability of a graphene cantilever against the attachment to an underlying conductor due to Casimir force. Research on graphene cantilevers picked up in recent years, fueled by the prospect of developing graphene-based sensors and nanoactuators, see, e.g., Refs. \cite{Li-APL2012,Miao-2014,AdP-2015,Martynov-APL2017,Miller-2017} and references therein. In this paper we show that besides being of practical interest, it can also provide an accurate method of measuring Casimir force. 

The system under consideration consists of a flat graphene sheet firmly attached to a flat solid surface, with a part of the sheet hanging above the surface of a conductor, see Fig.\ \ref{GC}. The width of the sheet in the y-direction is considered comparable to or greater than the length, $L$, of the hanging part in the x-direction, while $L$ is assumed to be large compared to the separation, $a$, from the conductor at $x = 0$. Casimir attraction to the conductor is provided by quantized vacuum and thermal fluctuations of the electromagnetic field between the conductor and the cantilever. The attraction results in the loss of stability of the cantilever against irreversibly bending down when $L$ exceedes some critical value $L_c$. Since the Casimir force increases with temperature, $L_c$ should depend on temperature. Inversely, for any fiixed length of the cantilever, $L$, which will be the case in experiment, there is a critical temperature, $T_c(L)$, above which the cantilever loses stability. 

Mechanical properties of the graphene cantilever depend on the number of atomic layers. As we shall see, even for a graphene monolayer, due to its strong flexural rigidity, the instability occurs when $L_c$ is large compared to $a$ and when the graphene sheet is only slightly bent towards the conductor. This allows one to use the proximity force approximation (PFA) \cite{Derjaguin-1958,Fosco-PRD2011}. Within the PFA one utilizes formulas derived for two parallel plates but treats the distance between the plates as a smooth slowly-varying function of coordinates. This approximation has been used to compute Casimir interaction between bodies of various geometries. It is considered reliable as long as the interacting surfaces are smooth, almost parallel, and close to each other. 

Generally speaking, Casimir's assumption that the electromagnetic radiation is reflected by the boundaries of the conductors is valid for frequencies below plasma frequencies. Similarly, Lifshitz theory \cite{Lifshitz,Derjaguin-1958}, that operates with macroscopic susceptibilities, breaks down at frequencies that exceed the absorption resonances of the material. Since the dominant contribution to the Casimir force comes from the photon wave vectors \cite{Lamoreaux} $k \sim 1/(4a)$, this translates into the lower limit on the separation between the two conducting surfaces, typically  $a \gtrsim 50$nm. The latter is in accordance with the fact that experiments performed down to $a \sim 0.1\mu$m agree well with theoretical formulas on the Casimir force. We will have it in mind when discussing the parameters of the cantilever problem. 

\section{The Model}

The energy of a 2D elastic memrane described by $z(x,y)$ is given by \cite{Nelson-JPhys1987,Katsnelson} 

\begin{equation}
H = \frac{1}{2}\int dxdy \left[ \kappa (\bm{\nabla}^2 z)^2 + \lambda u_{\alpha\alpha}^2 + 2 \mu u_{\alpha\beta}^2\right], %
\end{equation}

\noindent
where $\kappa(T)$ is the flexural stiffness constant, $\lambda$ and $\mu$ are Lam\'{e} elastic coefficients, ${\bf u}({\bf r})$ is the displacement field in the plane of the membrane, and
$u_{\alpha\beta} =\frac{1}{2}\left(\partial_{\alpha}u_{\beta} +\partial_{\beta}u_{\alpha} + \partial_{\alpha}z\partial_{\beta}z\right)$ is the strain tensor. For a suspended graphene sheet clipped at two edges (running, e.g., in the y-direction), the elastic strain terms in the energy are important even in the absence of phonons. Roughly speaking, they lead to \cite{Katsnelson,CZ-PRB2016} 
\begin{equation}\label{stretched}
H = \int dxdy \left[\frac{\kappa}{2}(\partial_x^{2}z)^{2}+\frac{\sigma}{2}(\partial_x z)^{2}\right]
\end{equation}
with $\sigma$ being the elastic stress applied to graphene. It is easy to see that for the typical values of the parameters of a monolayer graphene \cite{Jiang} stretched between two holders, $\kappa \sim 1$eV and $\sigma \sim 0.1 - 1$ J/m$^2$, the second term in Eq.\ (\ref{stretched}) dominates over the first term for any curvature radius of the graphene sheet in excess of $1$nm. This is similar to the case of an elastic string under tension \cite{LL}. On the contrary, the equilibrium mechanics of a graphene cantilever having a free end is determined by the first term in Eq.\ (\ref{stretched}), which makes the cantilever problem mathematically different from the problem of a suspended graphene sheet clipped at two edges.  

The energy of the Casimir attraction per unit area of a flat graphene sheet parallel to a flat surface of a conductor at a distance $a$  is given by \cite{Bordag} 
\begin{equation}\label{f}
f = -\beta \hbar c\left(\frac{1}{a^3} + \frac{1}{a_t a^2}\right),
\end{equation}
where 
\begin{equation}\label{beta}
\beta   =  \frac{ \alpha N}{128\pi}\left[\ln\left(1 + \frac{8}{\alpha N \pi}\right) + \frac{1}{2}\right] = 0.0003606235
\end{equation}
and
\begin{equation}\label{a_t}
a_{t}(T)   =  \frac{16 \pi \beta}{\xi(3)} \frac{\hbar c}{k_BT}=  0.01508\frac{\hbar c}{k_BT}
\end{equation}
with $\alpha = (4\pi\epsilon_0)^{-1}e^2/(\hbar c) = 1/137.036$ being the fine-structure constant, $N = 4$ being the number of fermion species for graphene, and $\zeta(3) = 1.20205$ being the value of Riemann zeta function $\zeta(x)$ at $x = 3$. The first term in Eq.\ (\ref{f}) is due to vacuum quantum fluctuations of the electromagnetic field, while the second term is due to thermal fluctuations. 

Within the PFA approximation $a$ in Eq.\ (\ref{f}) must be replaced with $z(x,y)$. Then the total energy of the graphene cantilever shown in Fig.\ \ref{GC} becomes
\begin{equation}
H_C = \int dxdy  \left[\frac{\kappa}{2}\left(\frac{d^2 z}{dx^2}\right)^2 - \frac{\beta \hbar c}{z^3} - \frac{\beta \hbar c}{a_t z^2}\right].
\end{equation}
It is convenient to introduce a characteristic length $b$ in the xy-plane and a characteristic energy $E_0$ according to
\begin{equation} 
b = \left(\frac{\kappa a^5}{3\beta \hbar c}\right)^{1/4}, \quad E_0 = \left(\frac{3\beta \kappa \hbar c }{a}\right)^{1/2}.
\end{equation}
For, e.g., $\kappa$ of order $1$eV and $a$ of order $1\mu$m the length $b$ is of order $10\mu$m and $E_0$ is of order $10$meV.  In terms of dimensionless variables
\begin{equation}
\bar{z} = \frac{z}{a}, \quad \bar{x} = \frac{x}{b},  \quad \bar{y} = \frac{y}{b}
\end{equation}
the energy becomes
\begin{equation}\label{dim-H}
\bar{H}_C \equiv \frac{H_C}{E_0} = \int d\bar{x}d\bar{y}  \left[\frac{1}{2}\left(\frac{d^2 \bar{z}}{d\bar{x}^2}\right)^2 - \frac{1}{3\bar{z}^3} - \frac{\theta}{2\bar{z}^2}\right].
\end{equation}
The only dependence on temperature is contained in the dimensionless parameter
\begin{equation}
\theta = \frac{2a}{3a_t} = \frac{T}{T_0}, \quad k_B T_0 = \frac{24 \pi \beta}{\xi(3)} \left(\frac{\hbar c}{a}\right) = 0.02262 \frac{\hbar c}{a}.
\end{equation}
At $a = 1\mu$m one obtains $T_0 = 51.8$K.

For $H[z(x)] = \int dx F[x, z(x), z'(x), z''(x), ..., z^{(k)}(x)]$ the Euler-Lagrange equation is
\begin{equation}
\frac{\partial F}{\partial z} - \frac{d}{dx}\frac{\partial F}{\partial z'} + \frac{d^2}{dx^2}\frac{\partial F}{\partial z''} - ... + (-1)^k \frac{d^k}{dx^k}\frac{\partial F}{\partial z^{(k)}} = 0.
\end{equation}
Eq.\ (\ref{dim-H}) then gives the following equilibrium equation for the graphene cantilever under Casimir force
\begin{equation}\label{ArbitraryT}
\frac{d^4\bar{z}}{d \bar{x}^4}  + \frac{1}{\bar{z}^4}  + \frac{\theta}{\bar{z}^3} = 0.
\end{equation}
It must be solved with the boundary conditions 
\begin{equation}\label{boundary}
\bar{z}(0) = 1, \quad \bar{z}'(0) = 0, \quad \bar{z}''(\bar{x} = l) = 0, \quad \bar{z}'''(\bar{x} = l) = 0,
\end{equation}
where 
\begin{equation}\label{l-def}
l = \frac{L}{b} = \left(\frac{3 \beta \hbar c}{\kappa a}\right)^{1/4} \frac{L}{a}.
\end{equation}
The last two boundary conditions correspond to the absence of force and torque at the free end of the cantilever respectively \cite{LL}.

\section{Stability of graphene cantilever}
The closer graphene is to the conductor the greater the Casimir force. It should be expected, therefore, that at some critical separation, or, equivalently, when $a$ is fixed, at some critical length, $L_c$, the graphene cantilever will lose its stability and its free end will sink down. Since at $\theta = 0$, neither Eq.\ (\ref{ArbitraryT}) nor the boundary conditions (\ref{boundary}) contain any dimensionless parameter, one should expect this to occur at $l \sim 1$. For, e.g., $a = 1\mu$m the condition $l =1$ gives $L = 8.4\mu$m. At finite temperature the effect of electromagnetic fluctuations must become stronger, resulting in the phase diagram, $T_c = T_c(a,L)$, separating stability region from the region where a part of the cantilever will attach to the conductor. In terms of dimensionless variables it corresponds to $\theta_c(l)$ that separates the two regions. However, since both $T_0$ and $l$ depend on $a$, the critical temperature $T_c$ depends in a non-trivial way on the separation from the conductor at $x = 0$ and on the length of the cantilever $L$. 

\begin{figure}[ht]
\begin{center}
\includegraphics[width=8.7cm,angle=0]{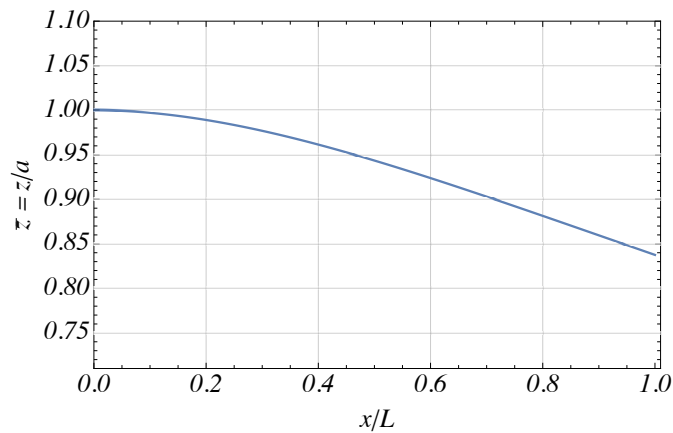}
\caption{Profile of a graphene cantilever attracted to an underlying conductor by the Casimir force at $T = 0$ and $l = 0.95 < l_{c0}$.}
\label{profile}
\end{center}
\end{figure}
Analytical solution of the forth-order equation (\ref{ArbitraryT}), either when it is dominated by the ${1}/{\bar{z}^4}$ term in the low temperature limit or by the ${\theta}/{\bar{z}^3}$ term in the high temperature limit, is unknown. Its numerical solution with the boundary conditions (\ref{boundary}) presents a challenge as it requires high precision. In order to find the critical length, $l_c$, we start at some small $l$ and increase it in small steps until the solution reveals instability. In our method the boundary value problem is approached as the initial-value problem with the boundary conditions treated as constraints. We begin with a small stable $l$ and increase it slowly to obtain the next solution using the initial values of the stable solution as the starting points. By formulating the problem in this way, we have been able to take the advantage of a shooting algorithm that is much faster and more accurate than the available boundary-value problem solvers for this type of highly nonlinear equation. The software used was Wolfram Mathematica. Most of the operations were done on a 40-core computing cluster.

Instability reveals itself in the emergence of the inflection point in the dependence of $\bar{z}$ on $l$ at the free end, as well as in the divergence of the second derivative of $\bar{z}(l)$. For each value of $l$ we obtained the value of $\theta$ above which a stable solution does not exist. We find that at  $T = 0$ the cantilever loses stability at $l_c \equiv l_{c0} \approx 0.984$ which is very close to our estimate $l_{c0} = 1$. At $T = 0$ and $ l = 0.95 < l_{c0}$ stable profile of a graphene cantilever attracted to an underlying conductor by the Casimir force is shown in Fig.\ \ref{profile}. 

\begin{figure}[ht]
\begin{center}
\includegraphics[width=8.7cm,angle=0]{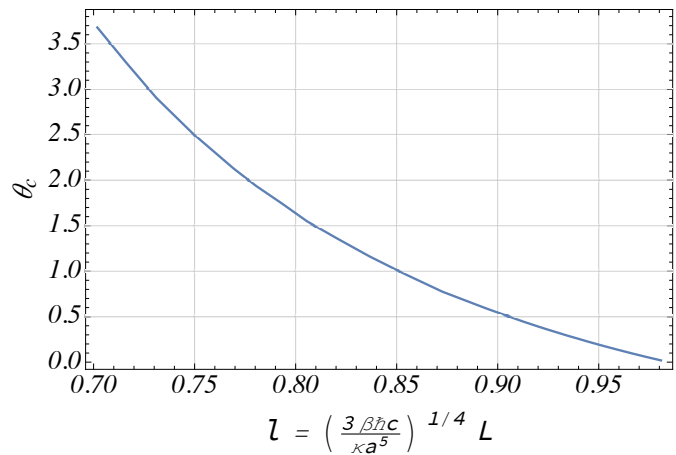}
\caption{Dependence of $\theta_c$ on $l$.}
\label{PD}
\end{center}
\end{figure}

The dependence of $\theta_c$ on $l$ in the experimentally accessible range $0.7 < l < l_{c0}$ is shown in Fig.\ \ref{PD}. It allows one to obtain the critical temperature $T_c = \theta T_0$ for any $\kappa$, $a$ and $L = b l$.

For, e.g., $\kappa = 1$eV, $a = 1\mu$m, and $L = 6\mu$m ($l = 0.72$), one obtains $T_c \approx 130$K. Choosing the appropriate value of the flexural stiffness $\kappa$ one can use our formulas and numerical results to obtain $T_c$ for a multilayer graphene. Already for a bilayer the value of $\kappa$ can be greater than for a monolayer graphene by as much as a factor $20$ \cite{Lindahl,Wei-NL2013}. Note, however, that due to the $1/4$-power dependence of $l$ on $\kappa$ in Eq.\ (\ref{l-def}) one should expect weak dependence of $T_c$ on $\kappa$. For a bilayer graphene  $T_c$ is likely to be greater than for a monolayer cantilever of the same length by no more than a factor $20^{1/4} \sim 2$. 

\section{Attachment to a conductor and exfoliation of graphene}
In considering the effect of the Casimir force on a graphene cantilever one has to make sure that it exceeds gravity, otherwise the latter must be incorporated into the problem. The corresponding condition at $T = 0$ is 
\begin{equation}
\frac{3\beta \hbar c}{a^4} \gg m_0 g \; \rightarrow \; a \ll \left(\frac{3 \beta \hbar c}{m_0 g}\right)^{1/4} \sim 30 \mu m,
\end{equation}
where we have used $m_0 = 3.8 \times 10^{-7}$kg/m$^2$ for the 2D mass density of a monolayer graphene. Thus, for $a \sim 1 \mu$m or less the force of gravity can be safely neglected in comparison with the Casimir force. While $m_0$ increases proportionally to the number of layers in a multilayer graphene, the $1/4$ power dependence of the critical $a$ on $\kappa$ provides a safety margin in this case too.

An interesting question is what happens above $T_c$ when graphene cantilever begins to bend irreversibly towards the conductor, eventually attaching to it. The initial dynamics of that process is described by the time-dependent equation 
\begin{equation}\label{dynamics}
\frac{\partial^2 \bar{z}}{\partial \bar{t}^2} + \frac{\partial^4\bar{z}}{\partial \bar{x}^4} + \frac{1}{\bar{z}^4} + \frac{\theta}{ \bar{z}^3} = 0
\end{equation}

\noindent
which is a generalization of Eq.\ (\ref{ArbitraryT}). Here 

\begin{equation}
\bar{t} = \frac{t}{t_0}, \quad t_0 = \left(\frac{m_0 a^5}{3\beta \hbar c}\right)^{1/2}.
\end{equation}

\begin{figure}[ht]
\begin{center}
\includegraphics[width=8.7cm,angle=0]{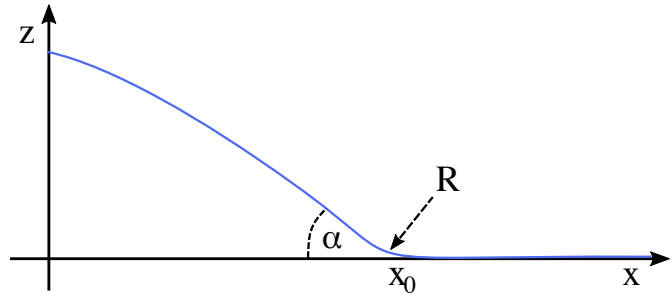}
\caption{Attachment profile of the graphene sheet.}
\label{exf}
\end{center}
\end{figure}

At $a = 1\mu$m the characteristic time $t_0$ is about $0.1$ms. Eq.\ (\ref{dynamics}) can be obtained by writing down the Lagrangian of the graphene cantilever subjected to the Casimir force:
\begin{equation}
{\cal{L}}_C = \int dxdy  \left[\frac{m_0}{2}\left(\frac{\partial z}{\partial t}\right)^2 - \frac{\kappa}{2}\left(\frac{\partial^2 z}{\partial x^2}\right)^2 + \frac{\beta \hbar c}{z^3} + \frac{\beta \hbar c}{a_t z^2}\right]
\end{equation}
The parameter that determines the scale of the cantilever frequency modes is
\begin{equation}
\omega_0 = \frac{1}{l^2t_0} = \frac{1}{L^2}\left(\frac{\kappa}{m_0}\right)^{1/2}.
\end{equation}
For  $\kappa \sim 1$eV and $L \sim 8\mu$m one obtains $\omega_0 \sim 10$kHz. 

Instability studied in the previous section can be also investigated by writing $\bar{z}(\bar{x},\bar{t}) = \bar{z}_{\rm eq}(\bar{x}) + \delta \bar{z}(\bar{x},\bar{t})$, linearizing Eq.\ (\ref{dynamics}) with respect to $\delta \bar{z}(\bar{x},\bar{t})$, and analyzing frequencies of small oscillations around the equilibrium static profile $\bar{z}_{\rm eq}(\bar{x})$. Instability occurs when the oscillation frequency develops an imaginary part on increasing $l$ above $l_c$ at a fixed $\theta$  or on increasing $\theta$ above $\theta_c$ at a fixed $l$. The speed of the corresponding dynamics that leads to the attachment is determined by the imaginary part of the frequency. On crossing $T_c$ the time for the cantilever to drop down must generally be of order $1/\omega_0$.

We shall now discuss the equilibrium profile of graphene attached to a conductor along the line $x = x_0$ in the xy-plane, see Fig.\ \ref{exf}. Interaction between the graphene sheet and the conductor is dominated by the Casimir force at large distances and by the Van der Waals force close to the attachment point. In the latter case the first-principle atomistic study \cite{Woods} is required. Nevertheless some useful relations can be obtained from the following consideration. Let us describe interaction by the function $U(z)$ that equals $f(z)$ of Eq.\ (\ref{f}) at large distances and provides a finite adhesion (wetting) energy per unit area, $U_0=U(0)$, at $z = 0$. The torque applied to the graphene sheet near the attachment point is \cite{LL} $\tau = \kappa({d^2 z}/{dx^2})L_y$, where $L_y$ is the size of the sheet in the $y$-direction. The work needed to raise graphene sheet of width $\delta x$ by $\delta z$ is
\begin{equation}
\tau \frac{d \delta z}{dx} L_y = \tau \frac{d^2 z}{dx^2} \delta x L_y = \kappa\left(\frac{d^2 z}{dx^2}\right)^2 \delta x L_y.
\end{equation}
It must be equated to the work,  $W = U(z) \delta x L_y$, required to change the interaction energy. This gives
\begin{equation}
\kappa\left(\frac{d^2 z}{dx^2}\right)^2 = U(z).
\end{equation}
Introducing the atomic-scale length $a_0 = \sqrt{{\kappa}/{U_0}}$ and dimensionless variables $\bar{x} = {x}/{a_0}, \bar{z} = {z}/{a_0}, u(\bar{z}) = {U(\bar{z})}/{U_0}$
one obtains $\left({d^2 \bar{z}}/{d\bar{x}^2}\right)^2 = u(\bar{z})$,
\begin{equation}\label{eq-u}
 \frac{d\bar{z}}{d\bar{x}} = -\sqrt{2 \int_0^{\bar{z}} d\bar{z} \sqrt{u(\bar{z})}},
\end{equation}
where the integration limits are chosen such that $d\bar{z}/d\bar{x} = 0$ at $\bar{z} = 0$, and the minus sign in front of the square root is chosen in accordance with Fig.\ \ref{exf}. 

At $\bar{z} \rightarrow 0$ one has ${d\bar{z}}/{d\bar{x}} = -\sqrt{2\bar{z}}$, which gives $\bar{z} = 0$ at $ \bar{x} > \bar{x}_0$ and  $\bar{z} = \frac{1}{2} (\bar{x} - \bar{x}_0)^2$ at $ \bar{x} < \bar{x}_0$. This implies that the curvature radius $R$ in Fig.\ \ref{exf} equals $a_0 = \sqrt{{\kappa}/{U_0}}$. In the opposite limit of $\bar{z} \gg 1$ Eq.\ (\ref{eq-u}) gives ${d\bar{z}}/{d\bar{x}} = -s$, where
\begin{equation}
s = \tan \alpha = \sqrt{2 \int_0^{\infty} d\bar{z} \sqrt{u(\bar{z})}} 
\end{equation}
is the slope of the graphene sheet on the approach to the attachment point $x = x_0$.  Note that the above equations must also describe the profile of the graphene sheet near the exfoliation point, see for review  Ref.\ \cite{Exf_mech}. Since the Casimir force rapidly decreases with increasing separation and $u(0) = 1$, the slope $s$ is dominated by the Van der Waals force and must be of order unity.

\section{Conclusions}

We have studied stability of graphene cantilever attracted to the underlying conductor by the Casimir force. At zero temperature the force is determined by vaccum fluctuations of the electromagnetic field. The critical length of the cantilever, $L_c$, depends on its distance from the conductor and flexural rigidity. At $T = 0$, for a monolayer graphene separated by $1\mu$m from the conductor, we obtained $L_c = L_{c0} \sim 8\mu$m. Casimir force increases as the temperature goes up due to the contribution of thermal photons. For a cantilever of length $L < L_{c0}$ there is a critical temperature $T_c$ above which the cantilever becomes unstable, with its free end falling and attaching to the conductor. $T_c$ as a function of the cantilever length has been computed. For, e.g., $L = 6\mu$m, we obtained $T_c \approx 130$K. 

Our results show that Casimir interaction should be taken into account in designing micro- and nano-electromechanical systems. Driven by the potential of graphene-based sensors and actuators the interest to graphene cantilevers has increased in recent years. Single-layer cantilevers have been manufactured. Enforcing such cantilevers with carbon nanotubes has been shown to significantly improve their vulnerability to structural defects and to increase their flexural stiffness \cite{Martynov-APL2017}. The effects studied in this paper are now within experimental reach with respect to the spatial scales and temperatures involved. They may provide a sensitive method for precision measurements of the Casimir force.

\section{Acknowledgements}
This work has been supported by the U.S. Department of Energy, Office of Science, under Grant No. DE-FG02-93ER45487.

\end{document}